\documentstyle[12pt]{article}
\begin{document}
\newcommand{\ket}[1] {\mbox{$ \vert #1 \rangle $}}
\newcommand{\bra}[1] {\mbox{$ \langle #1 \vert $}}
\newcommand{\bkn}[1] {\mbox{$ < #1 > $}}
\newcommand{\bk}[1] {\mbox{$ \langle #1 \rangle $}}
\newcommand{\scal}[2]{\mbox{$ \langle  #1 \vert #2 \rangle   $}}
\newcommand{\expect}[3] {\mbox{$ \bra{#1} #2 \ket{#3} $}}
\newcommand{\ki}{\mbox{$ \ket{\psi_i} $}}
\newcommand{\bi}{\mbox{$ \bra{\psi_i} $}}
\newcommand{\p} \prime
\newcommand{\e} \epsilon
\newcommand{\la} \lambda
\newcommand{\om} \omega   \newcommand{\Om} \Omega
\newcommand{\cc}{\mbox{$\cal C $}}
\newcommand{\w} {\hbox{ weak }}
\newcommand{\al} \alpha
\newcommand{\m} {\mbox{$\mu_k$}}
\newcommand{\n} {\mbox{$\nu_k$}}
\newcommand{\bt} \beta
\newcommand{\La} \Lambda 

\newcommand{\be} {\begin{equation}}
\newcommand{\ee} {\end{equation}}
\newcommand{\ba} {\begin{eqnarray}}
\newcommand{\ea} {\end{eqnarray}}

\def\lrD{\mathrel{{\cal D}\kern-1.em\raise1.75ex\hbox{$\leftrightarrow$}}}
\def\lr #1{\mathrel{#1\kern-1.25em\raise1.75ex\hbox{$\leftrightarrow$}}}
\newcommand{\lrpartial}[1] {\mbox{$\stackrel{\leftrightarrow}{\partial_{ #1}}$}}

\centerline{\LARGE\bf{
Particle creation and non-adiabatic
 }}\vskip .3 truecm
\centerline{\LARGE\bf{transitions in quantum cosmology}}
\vskip 1.5 truecm
\centerline{{\bf S. Massar}}
\centerline{Raymond and Beverly Sackler Faculty of Exact Sciences,}
\centerline{
School of Physics and Astronomy,
Tel-Aviv University, Tel-Aviv 69978, Israel}
\vskip 1.5 truecm
\centerline{{\bf R. Parentani}}
\centerline{Laboratoire de Math\'ematiques et Physique Th\'eorique,
CNRS UPRES A 6083,}
\centerline{Facult\' e des Sciences, 
Universit\'e de Tours, 37200 Tours, France. }
\vskip 1.5 truecm

\centerline{ MAY 1997}

\vskip 1.5 truecm
\centerline{\bf{abstract}}

The aim of this paper is to compute 
transitions amplitudes in quantum cosmology, and in particular
pair creation amplitudes and radiative transitions.
To this end, we apply a double adiabatic development to the
solutions of the Wheeler-DeWitt equation restricted to mini-superspace
wherein gravity is described by the scale factor $a$. 
The first development 
consists in working with instantaneous
eigenstates, in $a$, of the matter Hamiltonian. The second development
is applied to the gravitational part of the wave function and generalizes
the usual WKB approximation.
We then obtain an exact equation which replaces the Wheeler-DeWitt equation 
and determines the evolution, i.e. the dependence in $a$,
of the coefficients of this double expansion.
When working in the gravitational adiabatic approximation,
the simplified equation delivers the unitary evolution
of transition amplitudes occurring among instantaneous eigenstates.
Upon abandoning this 
approximation,
one finds that there is an additional coupling among matter states
living in expanding and contracting universes. Moreover one has to face 
also the Klein paradox, i.e. the generation of backward waves from 
an initially forward wave.  The interpretation
and the consequences of these unusual features are
only sketched in the present paper. 
Finally, the  examples of pair creation and radiative transitions are 
analyzed in detail to establish when and how
 the above mentioned unitary evolution coincides with
the Schr\" odinger evolution. 
\newpage

\section{Introduction}\label{int}

In General Relativity, 
one loses the usual Hamiltonian structure 
in which time parameterizes the evolution
since the theory is invariant under arbitrary
changes of coordinates.
In order to understand the consequences of this loss in quantum
cosmology, one must begin by reexamining the status of time in quantum
mechanics and quantum field theory. In particular, we draw the
attention to the distinction between the adiabatic
deformation of the states and non-adiabatic transitions
amongst states.
This distinction will be very useful
upon addressing the question of the evolution in quantum cosmology.

For simplicity,
we first consider the case
in which the Hamiltonian is time independent.
Then, its eigenstates are stationary and, in the absence
of interactions, the probability amplitudes to find the system in each
eigenstate are constant.
When a time independent interaction Hamiltonian is added,
its sole role is to displace
the energy levels from their values defined by the
free Hamiltonian to their dressed values defined
by the (stationary) eigenstates of the full Hamiltonian.

When the 
Hamiltonian is time dependent,
it has a double role.
First, it modifies {\it adiabatically} the energy levels defined
by its instantaneous eigenstates.
In the limit of slowly varying situations, one recovers
the former case in which the amplitudes stay constant.
 Secondly, 
it induces {\it non adiabatic transitions} between instantaneous 
eigenstates, a truly time 
dependent phenomenon since it characterizes quantum jumps. 
The importance of these transitions is controlled by $\partial_t
\Delta E / \Delta E^2$ where $\Delta E(t)$ is the energy difference
between the instantaneous eigenstates.
It goes without saying that as long as the interaction 
Hamiltonian is hermitian, 
 the existence of these transitions do not require
a revision of the interpretation of quantum mechanics since
the norm of the state stays constant.

In quantum field theory, 
time dependent gravitational fields induce
 both 
effects, although they are
somewhat hidden 
when one works -as is usually done- in a
Heisenberg picture. For instance for a free scalar field one would
proceed by computing the solutions of the Klein Gordon equation. Its
WKB solutions describe the adiabatic deformation
of positive frequency waves and lead to no transitions in Fock space. 
Instead, backscattering between positive and negative frequency
waves correspond to non-adiabatic transitions between states, i.e. pair
creation from vacuum. 
What plays the role of $\Delta E(t)$ in quantum mechanics is 
$2 \om(t)$, 
the energy of a pair.
Note that the existence of the backscattering amplitude leads to the
``Klein paradox'', namely the increase of the current 
carried by the forward wave 
when some backward wave is generated
(since the total current is conserved). This imposes that one abandon
a first quantized framework, since no conserved positive local
quantity can be built from these solutions.

In the first part of this article we review the notions of adiabatic
evolution and adiabatic transitions in a cosmological context in which
the 
evolution of the 
scale facor $a(t)$
is given from the outset. 
We show that $a$ itself can replace time and be used as the
parameter which controls both the adiabatic deformations of the
states and their non adiabatic transitions, although time still
appears in the relative phase accumulated by neighboring states. This
prepares for the analysis in quantum cosmology where 
$a$ is a dynamical variable.

We illustrate this formalism by two examples. The first is the
recovery of the Golden Rule, i.e. that transition probabilities grow
linearly with proper time lapses. We show how the residual 
time dependence mentioned above
guarantees that the transition rate is constant. 
(Thus a collection of such systems can
be viewed as a simple realization of a ``clock''.)
The second example, taken from field theory, is the process of pair
creation in cosmology. We show that pair creation acts are
described, in the Schr\" odinger picture, by non adiabatic
transitions between states having different instantaneous particle
number. Moreover, this second example introduces usefull tools to handle
the gravitational part of the solutions of the Wheeler-DeWitt
equation.

In the second part this article, we work in quantum cosmology. 
In that framework, the  physical states are solutions
of the Wheeler-DeWitt (WDW) equation.
Indeed, instead of having as before a Hamiltonian framework 
in which processes 
are parameterized in time, one has a constraint that expresses the
invariance of the theory under arbitrary changes of time 
(of time only since our analysis is restricted to mini-superspace)\cite{WDW}.
Therefore, to make contact with the conventional
analysis, one must understand how 
to define {adiabatic} deformations of matter states and
how to compute {non adiabatic} transition amplitudes without having a
physical time parameter at our disposal. 

The first point is easily addressed. Indeed, since the Wheeler-De Witt
equation is the sum of a kinetic term for gravity, a potential term
coming from the curvature of space, and the matter Hamiltonian, $H_M(a)$,
the adiabatic matter states are simply the 
instantaneous --in $a$-- eigenstates $H_M(a)$.
Therefore they coincide with
the instantaneous  eigenstates --in $t$-- defined in 
the Schr\" odinger framework, since in cosmology, 
the whole dependence in $t$ of $H_M$
comes through $a(t)$ only.

To address the second point, it is necessary
to carry out a second adiabatic expansion
applied to the gravitational part of the wave function. This generalizes the
usual WKB expansion which has been used in
\cite{banks}\cite{BV}\cite{vil}\cite{kiefer2} 
to extract a time parameter which governs the
Schr\" odinger equation in the mean geometry.
The central result of this paper is the (exact) equation
satisfied by the coefficients of this double adiabatic expansion
for both matter and gravity. 

When one works in the adiabatic approximation for gravity,
this equation has a very simple structure: it is a linear, first order 
equation in $\partial_a$ and bears a simple interpretation. 
Indeed, the identification of what replaces the probability
amplitudes in Schr\" odinger equation is straightforward.
It describes therefore
the non adiabatic transitions amplitudes
of matter parametrised by $a$ rather than $t$.
In this, we generalize \cite{wdwgf}--\cite{wdwin}
which were based on time dependent perturbation theory.
Indeed, in the adiabatic treatment, the dressing energy of matter states
is automatically included in the wave functions describing gravity
and non-adiabatic transitions are evaluated 
from gravitational wave functions wherein this 
back-reaction effect is fully taken into account.
Thus we no longer meet the problem\cite{wdwpt,wdwin} 
of defining transitions amplitudes that are non linear
in the coupling constant.
These advantages are illustrated in the last part of this article
when we return to the examples of radiative transitions and particle
creation now treated in quantum cosmology. We explicitly show when and 
how
one recovers the background field result.
Mathematically this requires to legitimize a first order expansion
in the energy difference $\Delta E(a)$ that characterizes the transition
upon investigation. Physically it means that the universe be 
macroscopic\cite{BV}\cite{vil}, i.e.
that the matter sources driving gravity be macroscopic.

When one abandons the adiabatic treatement for gravity,
one meets new effects which have no 
counterpart in quantum mechanics. These arise because
one must consider superpositions
of solutions describing contracting and expanding universes.
In this one inevitably encounters the ``Klein paradox'',
i.e. the generation of backscattered solutions
describing contracting universes when the wave was 
initialy purely forward. As in QFT, these backscattered solutions 
lead to an increase of the current carried by the forward wave
 describing expanding universes.
Our dynamical equation 
takes exactly into account the 
contribution of these backscattered solutions. 
It can thus be exploited
to analyze both conceptually and quantitatively their
consequences.
One must face indeed a double problem.
First, one must adopt a framework in which the generation of
backscattered waves can be
interpreted. In this respect, 
one must probably proceed to a ``third'' quantization.
Secondly, 
there is also the more pragmatic problem 
which consists in determining their consequences 
onto the matter propagation restricted
to the expanding sector, i.e. onto matter belonging to ``our'' universe. 

>From our exact dynamical equation, 
the much debated question of the unitarity of the evolution
can now be phrased in precise mathematical terms. 
One should indeed first identify what mathematically corresponds
to the probability amplitude to find a given adiabatic state at
a certain $a$. Only then can one hope to settle the question of the
unitarity of their evolution.
In this article we shall only sketch these problems. We hope to return
to them in a future publication.

We conclude by noting that the analysis of the Wheeler-De Witt 
equation presented here possesses illuminating analogues
in other physical systems. Indeed the basic problem, which is not
specific to quantum cosmology, is to enlarge the dynamics so that a
degree of freedom which was treated classicaly (the radius of the
universe $a(t)$) is treated quantum mechanicaly. Some of the
conceptual issues raised by this enlargement will of course be specific to
each
problem.

The first analogue is a uniformly
accelerated detector\cite{Unruh} which can make non
adiabatic transitions from its ground state to its excited state, and
{\it vice versa}. (The description of these events as non adiabatic
transitions can be found\cite{PBijpd}. The
treatement presented there is also closely related to our analysis 
of the Golden Rule). 
These transitions are characterized by a
temperature, in similar fashion that the pair creation events in
cosmology which we study in this article are characterized by a
temperature.
In these works the accelerated system follows a given classical
trajectory. However, in order to take into account recoil
effects one must go beyond this background field
approximation, and describe the accelerated system 
by WKB wave functions\cite{rec} or even by
second quantized fields\cite{su}. Then not only is the
trajectory treated quantum mechanically, but one
can also produce pairs of accelerated systems, which is strongly
reminiscent of third quantisation in cosmology. 
In both cases 
what should be interpretated as the probability amplitudes
to find a given adiabatic state in the 
forward sector do not obey a closed equation.
In other words, ``our'' expanding universe is not isolated.
The accelerated detector system also exhibits a
remarkable
thermodynamic consistency between the radiative transitions of the
detector and the pair production events. In quantum cosmology we will
also find an intimate connection between the dynamics of transitions
in the universe and the backscattering of universes. We have however
only exhibited this connection in this paper, and have not investigated
in detail its consequences.

The second analogue is the description of electronic non elasticity,
i.e. transitions between electronic states, during atomic or molecular
collisions\cite{LZ}\cite{S}. 
The simplest treatment is 
a background field approximation wherein the nuclear
trajectories are given classical functions $R(t)$. The electrons then
evolve according to the time dependent Hamiltonian
$H_{el}(R(t))$. (This corresponds of course, in our case, to QFT in a 
given space time.)
However one can go beyond this approximation and treat
the nuclei quantum mechanically 
(although in first quantization)\cite{DTK}\cite{MG}.
To this end one carries out a double
adiabatic development in the
electronic adiabatic states and the nuclear WKB wave functions. 
The equation for the coefficients of this double expansion describe
the non adiabatic transitions of the electrons, parametrised by the
nuclear coordinates in exactly the same way that the equation for
non adiabatic transitions of matter is recovered in quantum
cosmology. This equation can be used to
investigate the validity of the background field
approximation. In particular one can investigate in this context to
what extent the notion of a (temporal) succession
of electronic transitions depends on the nuclei being in a tight
wave packet. The analogy with quantum cosmology obtains therefore both
at the formal and at the conceptual level
(save for the interpretation of the backscattered waves).
Hopefully these systems could be used to investigate
experimentally some of the conceptual problems of quantum cosmology.

As a final comment, we wish to acknowledge the fact that many people
already appeal to the use of adiabaticity to investigate matter dynamics 
in quantum cosmology. Suprisingly, to our knowledge the full
power of the adiabatic treatement (by opposition to the adiabatic 
{\it approximation}) has not been exploited
previously. However,
after 
concluding this work,
we became aware of refs. \cite{embacher,mosta} 
wherein the full treatement is used, although the motivations of these
authors and the chosen exemples differ significantly from those of the
present paper. Indeed they have restricted their analysis to the ``Hawking
model'' wherein matter consists only of a massless scalar field
(because of its possible relevance to inflation), whereas we insist on
its generality and applicability to all matter processes. Furthermore they
have not used this formalism to analyse the origin of time in
quantum cosmology and the related conceptual issues 
which is the main focus of the present paper.

\section{The adiabatic treatment in quantum mechanics}\label{adia}

\subsection{Instantaneous eigenstates and non 
adiabatic transitions}\label{inst}

In this section we recall the basic
features of the adiabatic approximation applied to
quantum mechanics. We also review how to compute 
non-adiabatic transitions. To illustrate 
the usefulness of this formalism in a 
cosmological setting,
we apply it in subsections 2.2 and 2.3 to matter transitions 
and pair creation amplitudes.
We put special emphasis in the role played by the
expansion of the universe described by $a=a(t)$ since, 
in cosmology, the expansion 
is the cause of non adiabatic transitions.
Indeed, in the absence of expansion, the matter would be
described by stationary (un-interfering) eigenstates. 
This sets the stage for Section 4 where the same 
transitions will be described in quantum cosmology, wherein 
gravity is described by wave functions,
 solutions of the Wheeler-De Witt equation.

The Schr\" odinger equation that governs the evolution 
of the state $\ket{\psi(t)}$ is
\be
i \partial_t \ket{\psi(t)} = H(t) \ket{\psi(t)}
\label{sch}\ee
We expand $\ket{\psi(t)}$ in terms of the instantaneous eigenstates of 
the time dependent Hamiltonian $H(t)$:
\be
\ket{\psi(t)} = \sum_n \tilde c_n(t)\ket{\psi_n(t)}
\label{deceig}
\ee
where
\ba
H(t)\ket{\psi_n(t)} &=& E_n(t) \ket{\psi_n(t)}\nonumber\\
\scal{\psi_n(t)}{\psi_m(t)}&=& \delta_{n,m}
\label{insteig}\ea
Since $H(t)$ is hermitian, the 
eigen-energies $E_n(t)$ are real.
Inserting the decomposition eq. (\ref{deceig}) into 
eq. (\ref{sch}) yields the equation for the
$\tilde c_n(t)$
\be
i\partial_t \tilde c_n(t) = E_n(t) \ \tilde c_n (t) + i
\sum_{m} \scal{\partial_t\psi_m(t)}{ \psi_n(t)} \ \tilde c_m(t)= \sum_m
H^{eff}_{n m }(t)\ \tilde c_m(t)
\label{eqcn}
\ee 
where we have introduced, for later convenience,
 the effective Hamiltonian $H^{eff}(t)$.
Note that the antisymmetry of   
$\scal{\partial_t \psi_m(t)}{\psi_n(t)} = - 
\scal{\psi_m(t)}{\partial_t\psi_n(t)}$ 
and the reality of $E_n(t)$ ensure that $H^{eff}(t)$ is hermitian.
Therefore the 
conservation of probability $\sum_n \vert \tilde c_n(t) \vert^2 = 1$
is guaranteed.

Since the states $\ket{\psi_n(t)}$ are defined up to an arbitrary phase,
we may use this freedom to require
\be
\scal{\psi_n(t)}{\partial_t\psi_n(t)}=0
\ee
Note however that if $\ket{\psi_n(t)}$ evolves
cyclically during the interval $t_0 , t_1$, then 
the state at $t_1$ will in general differ from
state at time $t_0$ by a phase, known as the Berry phase\cite{Berry}.
Berry's phase 
will not be of concern to us in the present paper
since in mini-superspace the matter states depend only on
{\em one}
parameter, the scale factor $a$. 
In this case Berry's phase vanishes, since all paths can be mapped one
onto another by  a reparameterisation. However if one goes beyond mini
super space, one should take the Berry phase into account.

Upon absorbing, as in time dependent
perturbation theory, the diagonal part of $H^{eff}(t)$ into 
the definition of new coefficients
$ c_n(t)= \tilde c_n(t) e^{i\int^t dt E_n(t)}$, one obtains
\be
\partial_t c_n = \sum
_{m\neq n} 
\scal{\partial_t\psi_m(t)}{ \psi_n(t)}\ e^{-i\int^t dt'
( E_m(t') - E_n(t') )}\ c_m(t)
\label{eqtcn}
\ee 
At this point, it is important to notice that this
equation is ``almost'' reparametrisation invariant.
The only term that does not satisfy this invariance is the 
phase containing the change in energy $E_m(t') - E_n(t')$.
As we shall show in section 2.2, it is this factor that guarantees that 
the Golden Rule is recovered, i.e. that probabilities
increase linearly with respect to (proper) time.
In a cosmological situation,
the sole time dependence of $H(t)$
comes through the expansion law $a=a(t)$.
Therefore the eigenstates $\ket{\psi_n(a)}$
and the eigenvalues $E_n(a)$ depend on $t$ through $a(t)$
only, by virtue of their definition, see eq. (\ref{insteig}). 
Thus, one can write eq. (\ref{eqtcn}) directly in terms of $a$
as
\be
\partial_a c_n = \sum
_{m\neq n} 
\scal{\partial_a\psi_m(a)}{ \psi_n(a)}\ 
e^{-i\int^a da' (dt/da')(E_m(a') - E_n(a'))}\ c_m(a)
\label{eqtcn2}
\ee
We shall see in Section 3 that it is in that guise, i.e. wherein $t$ appears
only as a factor of $E_m(a) - E_n(a)$ and in the form $dt/da$,
that the evolution of the $c_n$ will be delivered in quantum cosmology.

To proceed we must evaluate $\scal{ \psi_n(t)}{\partial_t \psi_m(t)}$
($n\neq m$).
If  $H$ is non degenerate
this is given directly in terms of the
Hamiltonian. Indeed  differentiating eq. (\ref{insteig})  
\be
\partial_t H \ket{\psi_n} + H \ket{ \partial_t \psi_n}
= \partial_t E_n  \ket{ \psi_n} + E_n(t) \ket{ \partial_t \psi_n}
\ee
and taking the scalar product with $\bra{\psi_m}$ yields
\ba
\scal{\psi_m}{\partial_t\psi_n}
&=&
{\bra{\psi_m} \partial_t H \ket{\psi_n}\over E_n - E_m }\quad , \quad n \neq m
\label{dHdt}
\ea
Thus the time dependence of $H(t)$
leads to non-vanishing off
diagonal elements in $H^{eff}$. 
These have two physical consequences. First
they modify the instantaneous eigenstates and 
eigenvalues\cite{Berry2,Jackiw}. Secondly, they
induce 
transitions among eigenstates\cite{Messiah}.

The first effect is 
described by carrying out an
instantaneous diagonalisation of $H^{eff}(t)$
to obtain new eigenstates and eigenvalues:
$H^{eff}(t)\ket{\psi_n^{(1)}(t)} = E_n^{(1)}(t) \ket{\psi_n^{(1)}(t)}$. 
After having diagonalised $H^{eff}(t)$, one can decompose $\ket{\psi}$ as
$\ket{\psi(t)}= \sum_n \tilde c_n^{(1)}(t) \ket{\psi_n^{(1)}(t)}$. 
The evolution 
of the new coefficients 
$ \tilde c_n^{(1)}(t)$ is given by a new effective Hamiltonian
$H^{eff(1)}(t)$. One can then iterate the procedure and diagonalise 
$H^{eff(1)}(t)$. This generates an asymptotic series for the state
$\ket{\psi(t)}$. The series starts diverging when the difference
between successive terms becomes comparable to the amplitude for non
adiabatic transition. 

To work in the adiabatic approximation consists in neglecting
the second effect, namely non adiabatic transitions 
between instantaneous eigenstates.
In general, to compute the amplitude of these transitions, 
it is sufficient to consider only a couple of 
states, say $\ket{\psi_1}$ and $\ket{\psi_2}$.
Then eq. (\ref{eqtcn}) reduces to
\ba
\partial_t  c_1 &=& \bk{\partial_t \psi_2| \psi_1} \ e^{-i\int^t
dt' (E_2 - E_1)} \ c_2 \nonumber\\
\partial_t  c_2 &=&\bk{ \partial_t \psi_1| \psi_2}\ e^{-i\int^t
dt' (E_1 - E_2)} \ c_1
\label{twoeq}
\ea
Let us further assume that at $t=-\infty$ the system is in the state
$\psi_1$, i.e. $c_1 (-\infty) =1$, $c_2 (-\infty) =0$.
Then the 
transition amplitude is given by $c_2(+\infty)$. 
To evaluate
it, we proceed exactly like in perturbation theory 
by taking $c_1 =1$ 
in eq. (\ref{twoeq}) to obtain
\be
c_2(+\infty) \simeq
\int_{-\infty}^{+\infty} dt 
\bk{\psi_2|\partial_t \psi_1}\  e^{-i\int^t_{-\infty}
dt' (E_1(t') - E_2(t'))}
\ee
We consider only the case in which the resonance condition
$E_1(t) - E_2(t) =0$ has no real solution.
(This is generally the case in physical problems, because if their is
an interaction between the two energy levels the degeneracy is lifted to give rise
to an avoided crossing\cite{Messiah}).
Then, when the transition amplitude is small,
it can be correctly evaluated by 
a saddle point treatement, exactly like in other cases of
classically forbidden transitions such as tunneling processes. 
The saddle is located at the {\it complex} value
of $t^*$, solution of $E_1(t^*) - E_2(t^*) =0$. The result is thus
\be
c_2(+\infty) \simeq C\ 
 e^{-i\int^{t^*}_{-\infty}
dt' (E_1 - E_2)}
\label{correct}\ee
A rigorous analysis  
shows that $C$ tends to $1$ in the adiabatic limit,
$\partial_t (E_1 - E_2)/ (E_1 - E_2)^2 \to 0$, see \cite{DDP}. 
If the problem cannot be reduced to that of two coupled states, then
the analysis is more complicated because the non adiabatic transitions
can occur by ``hopping'' from one energy eigenstate to the other, 
see \cite{HwangP}.

\subsection{The Golden Rule and non adiabatic transitions}\label{gold}

In this section we apply the adiabatic formalism 
to radiative transitions in
cosmology in order to display the crucial role
played by the classical expansion $a=a(t)$.
The particular problem we consider is the
disintegration of a heavy state of mass $M$ into a light state of mass $m$
and a conformally coupled scalar quantum $\phi$. 
In the language of Unruh\cite{Unruh}, $M$ and $m$ represent the excited and
ground state of a particle detector of the $\phi$ field.
As in the Unruh analysis, the recoil of the heavy system is neglected.
Because this transition takes place in an expanding universe,
the energy of the emitted quantum (measured in proper time)
is $k / a(t)$ where $k^2$ is the conserved eigenvalue of the
three Laplacian.
The free initial and final energies are thus 
$E^0_{initial} = M$ and $E^0_{final}=m + k/a$.
They coincide when the Doppler red shifted energy of the photon
 is equal to $M-m$, i.e. when
\be a(t_0) = a_0 ={ k \over \Delta M }\label{cross}\ee
 Thus, in time dependent perturbation theory based on free states,  
the transition 
will occur around that value of $a$ wherein the energy of the photon 
resonates with $\Delta M$\cite{wdwpt}.
In what follows, we shall see that
 non adiabatic transition will also occur in the vicinity of $t_0$
even though the meaning of ``to make a transition'' will change. 

The unperturbed states are coupled by an interaction Hamiltonian
\be
H_{int}(a) = g \left( |M><m| \hat \phi + h.c.\right)
= g \left( |M><m| \sum_k {1 \over a k^{1/2}} (b_k + b^\dagger_k) + h.c.\right)
\ee
Thus the full Hamiltonian 
is
\ba
H(a) &=& \left( \begin{array}{cc}
M & {g / a k^{1/2}} \\
{g / a k^{1/2}} & m + {k / a} \end{array}\right)
\ea
As pointed after eq. (\ref{eqtcn}), the time dependence
comes only through $a= a(t)$. Therefore, one can directly work with
$a$ as relevant parameter.
Because of the non-diagonal elements 
induced by $H_{int}$, the degeneracy of the
free states at $a_0$ is
lifted to give an ``avoided crossing''\cite{Messiah}
when one considers the instantaneous 
eigenstates, $\ket{\psi_+}$ and $\ket{\psi_-}$,
which diagonalize $H(a)$.
Indeed, their corresponding 
eigenvalues are
 \be
E_\pm(a) =
{M + m + k/a \over 2} \pm \sqrt{(M - m -k/a)^2/4 + g^2/a^2k}
\label{valuu}
\ee
They coincide only for the complex values 
\be
a^*_{\pm} = a_0  \pm {i 2 g\over  \Delta M k^{1/2}} = a_0 \pm i \Delta a_i 
\label{compa}
\ee
The principle effect of this avoided crossing (for real values of $a$)
is that the instantaneous eigenstate $\ket{\psi_+}$ ($\ket{\psi_-}$)
which for $a \ll a_0$ coincides with the free initial (final) state 
$\ket{\psi_{inital}}$ ($\ket{\psi_{final}}$)
tends to the free 
state $\ket{\psi_{final}}$ ($\ket{\psi_{initial}}$) for $ a \gg a_0$.
Thus what corresponds to
the transition amplitude in terms of free states 
is given by the amplitude to stay in the
same instantaneous eigenstate, whereas the amplitude to stay in the
same free state is given by the amplitude to make a  non-adiabatic
transition from $\ket{\psi_+}$ to $\ket{\psi_-}$.

By writing the probability to make a non-adiabatic transition, 
that is to stay in the excited state $M$ as
\ba
P_{no\ transition}(k)= e^{-2 {\cal I}_k}
\label{Pnot}\ea
using 
eqs. (\ref{eqtcn2}, \ref{correct}), we have, 
\be
{\cal I}_k =  Im 
\int_{a_0} ^{a^*_{+}}
\!da \; { dt \over da}(a)
\left[E_+(a) - E_-(a)
\right]
\label{integralpos}
\ee
since only the integral for imaginary (positive) values of $a$ contributes 
to ${\cal I}_k$.
To evaluate this integral, we assume that $g$ be small enough so as
to justify the following approximation
\ba
{\cal I}_k  &\simeq& {1 \over a_0}
{ dt \over da}(a_0) Im
\int_{a_0} ^ {a^*} da \sqrt{(\Delta M a -k)^2 + 4g^2/k}
=   { \pi g^2 \over k^2} { dt \over da}(a_0)
\label{Pk}
\ea
Within this quadratic approximation, 
the probability for no transition is given {\it exactly} by
eq. (\ref{Pnot}), see \cite{S}. This exact result plays an important role
here since it guarantees that the probability not to decay decreases
exponentially in (proper) time, see eq. (\ref{rate}).
This probability is given by 
\ba
P_{no\ transition}(a)  =  \prod_{k(a)} P_{no\ transition}(k)
= e^{- 2 \int^{k(a)} {d^3 k \over (2 \pi)^3} { \cal I}_k}
\ea
To carry out the integral over $k$ in the exponent, we first
 use the isotropy of the space-time to write $d^3 k = 4 \pi k^2
dk$ (for simplicity, we work with flat three-geometries). Furthermore,  
 by using the resonance condition eq. (\ref{cross}), we can write $dk $ as 
$ \Delta M da_0$ and thus relate the final value of $a$ to a maximum
value of $k$, see \cite{GO} for a more detailed treatment
concerning the relationships between the integration over momenta 
and an integration over time (or $a$). Using that relation and eq. (\ref{Pk}), 
one gets
\be
 P_{no\ transition}(a) = 
e^{ - {  g^2 \Delta M  \over \pi} \int^a d a_0  { dt \over
da} (a_0)} = e^{ - {  g^2 \Delta M  \over \pi}\int^{t(a)} dt}
\label{rate}
\ee
corresponding to an exponential decay probability in proper time. 
It agrees with the probability evaluated using more traditional 
techniques\cite{wdwpt}.

It is interesting that in the present formalism the role of time and
energy seem to be reversed compared to usual treatment of the  Golden
Rule. Namely if the particle does not decay it seems to be propagating
in complex time, see eq. (\ref{compa}), whereas one usually ascribes
a complex energy
to unstable states. And the time $T=\int dt$ which
multiplies the decay rate in eq. (\ref{rate}) arises from an integral
over $k$, whereas it is usually obtained by letting the excited states
interact with the field through $H_{int}$ for a finite time $T$.

The treatment presented here can be
enlarged to include the effects of momentum recoils. This 
gives rise, as discussed in
\cite{wdwpt}, to the concept of spatial displacement.

\subsection{Particle creation as a non adiabatic process}\label{part}

In this section we apply the adiabatic formalism to particle creation in the
universe. 
This particle production\cite{Parker,BD} is a generic feature of quantum field
theory which is due to 
the expansion law $a=a(t)$. 

When 
the cosmological expansion is described 
by a classical law, the equation for free matter fields is linear
in the dynamical variables.
Therefore, it is particularly useful to work
in the Heisenberg picture since probability amplitudes to create 
pairs can be directly evaluated from the sole knowledge of the Bogoliubov
coefficients. However, upon abandoning the   
background field approximation for gravity by working in quantum cosmology,
one looses the linearity of the field equation since $a$ 
is now a dynamical variable coupled to matter at a quantum level.
As noted in \cite{wdwgf}, this loss of linearity imposes to 
work
in a Schr\" odinger picture since every transition amplitude
should be computed separately. Therefore, in what follows,
as a preparation for the analysis in quantum cosmology,
we study pair creation amplitudes in the Schr\" odinger picture.
In this picture, they
correspond to non-adiabatic transitions
among instantaneous eigen-states.

We shall work in a closed Robertson-Walker universe
\ba
ds^2 &=& -dt^2 + a^2(t) \sum_{i,j} h_{ij}dx^idx^j
\label{1}\ea
where 
\be
h_{ij}= {dr^2 \over 1 - r^2} + r^2 d\theta^2 + r^2sin^2\theta d\phi^2
\label{2}\ee
The action for a massive hermitian field is
\ba
L &=& {1 \over 2}\int dt \int d^3x \sqrt{h}a^3 \left[ \partial_t
  \phi\partial_t \phi - a^{-2} h^{ij}\partial_i \phi \partial_j \phi -
  m^2 \phi^2\right]
\label{3}\ea
The symmetry of 
the space allows 
$\phi$ to be decomposed as 
\be
\phi (t,x) = \sum_k{\cal Y}_k (x) \phi_k(t) 
\label{4}\ee
where ${\cal Y}_k$ are normalized eigenfunctions of the spatial Laplacian.
Using eq. (\ref{4}), the action reads
\be
L = \sum_k \int\!dt  a^3(t) \left[  \partial_t \phi_k \partial_t \phi_{-k}
- \Omega_k^2(t)
\phi_k\phi_{-k}\right] 
\label{7}\ee
where $\Omega_k(a)=(m^2 + k^2 a^{-2})^{1/2}$ is the time
dependent frequency at fixed $k$. 

The momentum conjugate to $\phi_k$ is $\Pi_k= a^3 \partial_t
\phi_{-k}$, and the Hamiltonian is
\ba
H(a) = \sum_k  \left[ a^{-3} \Pi_k \Pi_{-k} + a^3
\Omega_k^2(a)\phi_k \phi_{-k}\right] = \sum_k H_k(a)
\label{Ham9}\ea

We now carry out the instantaneous  diagonalisation of $H_k(a)$. 
To this end we express the
Schr\" odinger operators $\phi_k$ and $\Pi_k$ in terms of the $a$-dependent 
creation and destruction operators $b_k(a)$ 
\ba
\phi_k &=& {1 \over \sqrt{2} (a^3 \Omega_k )^{1/2} } (b_k + b_{-k}^\dagger)
\nonumber\\
\Pi_k &=& {1 \over \sqrt{2}}{i 
(a^3 \Omega_k )^{1/2}
} (b_{-k} - b_k^\dagger)
\label{deca}
\ea
Inserting eq. (\ref{deca}) into 
eq. (\ref{Ham9}) yields
\be
H_k(a) = \Omega_k(a) (b_k^\dagger b_k + b_{-k}^\dagger b_{-k})
\label{Hdiag}\ee
where we have subtracted the ground state energy by normal ordering
with respect to the instantaneous operators $ b_k$.
The instantaneous eigenstates are therefore those of definite particle number
\ba
\ket{\psi_{\{n_k\}}(a)} &=& \prod_k 
{(b_k^\dagger)^{n_k} \over \sqrt{n_k!}}\ket{0_a}\nonumber\\
E_{\{n_k\}}(a) &=& \sum_k \Omega_k (a)  n_k
\label{instaa}
\ea
where $\{n_k\}$ denotes the collection of occupation numbers and
$\ket{0_a}$ the instantaneous vacuum state.

The equation governing non adiabatic transitions 
is
\ba
\partial_a  c_{\{n_k\}}
&=&\sum_{\{n_{k'}\}}
\scal{\partial_a \psi_{\{n_{k'}\}}}{\psi_{\{n_k\}}}
e^{-i \int^{t(a)}\! dt'\ 
( E_{\{n_{k'}\}} - E_{\{n_k\}})}c_{\{n_{k'}\}}
\ea
where the state of the $\phi$ field has been decomposed
as
\ba
\ket{\psi(a)}&=&\sum_{\{n_k\}} c_{\{n_k\}}(a) \;
\ket{\psi_{\{n_k\}}(a)}
\label{decomposeas}
\ea
To obtain an expression for 
$\scal{\psi_{\{n_k\}}}{\partial_a \psi_{\{n_{k'}\}}}$, 
we must calculate $\partial_a H_k$, see eq. (\ref{dHdt}),
\be
\partial_a H_k = \partial_a  \Omega_k 
\left( b_k^\dagger b_k + b_{-k}^\dagger b_{-k} \right)
+ \Omega_k (a) { \partial_a  (a^3 \Omega_k) \over a^3 \Omega_k}
\left( b_k^\dagger b_{-k}^\dagger  + b_{k} b_{-k}\right)
\ee 
$\partial_a H_k$ contains two terms. The first is diagonal in the
basis of instantaneous eigenstates and is due to the adiabatic deformation of the
eigen-energy through 
$\Omega_k (a)$.
The second term is due to the time dependence of the operators
$b_k(a)$, $ b_k^\dagger(a)$. It couples states which differ by a pair
of particles of opposite momentum and engenders non adiabatic 
transitions. 
Thus the corresponding physical
processes 
are pair creation (or destruction) of particles with opposite momentum.

We now consider the process of creation of a pair from vacuum. Thus
we seek the amplitude for the non adiabatic transition from the ground
state $\ket{0_a}$ to the two particle state $b_k^\dagger b_{-k}^\dagger
\ket{0_a} = \ket{1_k, 1_{-k}}$. We assume that the creation process
be small enough so as to legitimate a description restricted to these
two states. In that case,
the equations are, see eq. (\ref{twoeq}),
\ba
\partial_a c_0 &=&  {1 \over2}{ \partial_a  (a^3 \Omega_k) \over a^3 \Omega_k}
e^{-2 i \int^{t(a)}\! dt'  \Omega_k} c_{k,-k}\nonumber\\
\partial_a c_{k,-k} &=&  {1 \over2}   
{ \partial_a  (a^3 \Omega_k) \over a^3 \Omega_k}
e^{2i \int^{t(a)}\! dt' \Omega_k}
c_{0}
\ea
Thus, for rare creation acts, the probability to produce a pair is, 
see eqs. (\ref{correct}), 
\be
\vert A_{pair}\vert^2
= e^{-Im \left[4 \int_{-\infty}^{t(a^*)} \sqrt{m^2 + k^2/ a^2} dt \right]}
\label{paircreat2}
\ee
where $a^*$ is the complex saddle point defined by $\Omega_k(a^*)
=0$ in the positive imaginary half plane.

To illustrate the above formalism, we apply it to
de Sitter space. 
In that case, the expansion is given by
\be
 a(t) = {1 \over \sqrt{\Lambda}} cosh \sqrt{\Lambda} t
\ee
Thus the complex saddle point of eq. (\ref{paircreat2}) is 
\ba
a^*_\pm &=& \pm i k/m\nonumber\\
t^*_{\pm}&=& t_0 \pm i \Delta t_i = {1 \over \sqrt{\La}} arcsinh {m \over
  k \sqrt {\La}} \pm i \pi / 2 \sqrt{\La} 
\label{tstar}\ea 
It lies in the vicinity of the
real time $t_0$
at which $k/a \simeq m$, i.e. where the particle
 ceases to be relativistic.
When $m> \sqrt{\La}$, the probability to produce a pair is small
and eq. (\ref{paircreat2}) can be used.
The integral 
is most easily evaluated in the 
limit of large $k$, i.e. for $\vert a^* \vert \gg 1/m$.
Indeed, in that limit, as in eq. (\ref{Pk}), the integrant is quadratic in $a$
\ba
Im \left[ 4 \int_{-\infty}^{t^*} \sqrt{m^2 + k^2/ a^2} dt \right] &=&
Im \left[ 4 \int_{a_0}^{a^*} da {(dt/da) \over a}  \sqrt{m^2 a^2+ k^2} \right]
\nonumber\\
&\simeq &
Im \left[ 4 { 1 \over a_0^2 \sqrt{\Lambda}} \int_{a_0}^{a^*} da \sqrt{m^2 a^2+ k^2}\right] 
= \pi m / \sqrt{\Lambda}
\ea
where we have used $da/dt = a \sqrt{\Lambda}$ valid for large $a$, and replaced
$a_0= a(t_0)=k/m$.
Thus we have
\be\vert A_{pair}\vert^2
\simeq e^{ -\pi m / \sqrt{\Lambda}}.
\label{PairdS}\ee
This 
corresponds to the Boltzmannian tail of a thermal distribution at the
de Sitter temperature $T= \sqrt{\Lambda}/ \pi$. 
This is as it should be since
the use of the saddle point
calculation and neglection of multi pair states cannot reproduce the
full Bose Einstein distribution\cite{BD}.

Before considering how to recover
these transitions in quantum cosmology,
we shall first relate the present analysis to the more conventional analysis
of pair creation
based on the 
solutions of the Klein Gordon
equation. The main motivation for this study is that it introduces 
tools that will be extremely usefull upon studying the solutions
of the WDW equation.

\subsection{The WKB approximation and pair creation}

The aim of this section is to phrase in adiabatic terms
the standard results concerning
the WKB approximation and the backscattering amplitude
 above a barrier. This will clarify
the relationship between pair creation probabilities as calculated
above
and backscattering.
The necessity of normalizing properly the Wronskian
of the WKB solutions will be stressed.
This will also prepare for the analysis of the solutions of the WDW
equation for which a correct normalisation will also be essential.
 

Consider 
a one dimensional harmonic oscillator whose 
frequency depends on time
\be
\left[ \partial^2_t + \omega^2(t) \right]\varphi(t) =0
\label{7bis}\ee
The WKB solution of unit Wronskian is
\ba
\chi(t) = {e^{-i\int^t_{-\infty} dt \omega(t)} \over \sqrt{2 \omega(t)}
}
\ea
One can then expand the exact solution in terms of $\chi(t)$ and $\chi(t)^*$
as
\be\varphi(t) = c(t) \chi(t) + d(t)\chi^*(t)\label{wkbexp}\ee 
We determine completely  
the coefficients $c$ and $d$ by requiring that
\be
i \partial_t \varphi(t) =\omega(t) \left[ c(t) \chi(t) - d(t)\chi^*(t)\right]
\label{OTHER}\ee
This guarantees that $c$ and $d$ are constant in the WKB limit, i.e. in the
adiabatic limit $\partial_t \omega(t)/ \omega(t)^2 \to 0$.
Moreover,
the conserved current (or Wronskian)
takes the simple form
\be
\varphi^* i\lrpartial{t} \varphi = \vert c(t) \vert^2 - \vert d(t)  \vert^2 
=constant
\ee
This equation replaces the unitary relation governing the coefficients
$c_n(t)$, solutions of the
Schr\" odinger equation, see 2.1. 
To obtain $c$ and $d$, one
 differentiates eq. (\ref{wkbexp}) and compares
the result with 
eq. (\ref{OTHER}), so as to obtain 
\be
\chi \partial_t c + \chi^* \partial_t d -{1\over 2} {\partial_t \omega \over \omega}
(\chi c + \chi^* d)=0
\label{rel}\ee
Now differentiating eq. (\ref{OTHER}), inserting the result into 
eq. (\ref{7bis}),
and using eq. (\ref{rel}) to eliminate either $\partial_t c$ or 
$\partial_t d$ yields the coupled first order equations 
\ba
\partial_t c &=&
{1\over 2} {\partial_t \omega \over \omega}e^{-2i\int^t_{-\infty} dt \omega}d
\nonumber\\
\partial_t d &=&
{1\over 2} {\partial_t \omega \over \omega}e^{+2i\int^t dt \omega}c
\label{1WKB}
\ea
These equations are equivalent to the original equation for $\varphi$,
eq. (\ref{7bis}). This has been discussed in detail in \cite{Baird}.
They constitute a convenient starting 
point for the evaluation of the non-adiabatic transitions, which
in the present case  correspond to backscattering.
As in Section 2.1 they are given by 
the value of value of $d$ at late times, given that $c=1,d =0$ at 
early times. 
Using the same type of analysis as in section 2.1,
one obtains
\be
d(+\infty)\simeq 
\int_{-\infty}^{+\infty} dt {\partial_t \omega \over 2 \omega} 
e^{-2i \int^t_{-\infty} \omega(t') dt'}
\ee
As before, the saddle point is at the complex value 
$t^*$ such that $\omega(t^*)=0$.  
The norm of $d$ is thus
\be
\vert d(+\infty) \vert^2 \simeq C \ e^{ - 4  Im
\int^{t^*}_{-\infty} 
\omega(t') dt' }\label{assB}
\ee
A rigorous analysis shows that $C$ tends 
to $1$ in the adiabatic limit\cite{RBS}. 


To obtain the relation between this norm and the {\it probability}
to produce a pair of quanta as calculated in section 2.3,
we must first determine the frequency $\omega(t)$ that corresponds
to $\Omega_k(a)$ which caracterizes the instantaneous energy of a 
particle of momentum $k$. A standart calculation yields
\be
\omega(t) = \Omega_k(a(t)) -\left( {3 \over 4} { 
(\partial_t a)^2 \over a^2} + {3 \over 2} {\partial^2_t a \over a}
\right)
\ee
The additional terms arise from the factor of $a^{3/2}$ which relates
the amplitude of 
$\phi_k(t)$ appearing in eq. (\ref{Ham9})
to 
$\varphi(t)$.
It is important to note that the backscattering amplitude and the pair
creation amplitude are related by
$\vert d(+\infty) \vert^2=
\vert A_{pair}\vert^2$ only if 
the Wronskians of the WKB waves that multiply
$c$ and $d$ are chosen in conformity with the second quantized framework,
i.e. are equal to one. 
In the present case of a single linear equation, only the
relative normalisation plays a role, however upon introducing
coupling among many oscillators, the normalisation is completely
 fixed, see \cite{wdwgf,wdwpt,wdwin,rec,su}. 

We wish to emphasize this last point since it will play a crucial role
in quantum cosmology.
The question of the choice of the Wronskians can be considered from two
different perspectives. The usual approach is based on the second quantized
framework. In that case, the choice of unit Wronskian is directly dictated by
the particle interpretation of the Heisenberg creation and
destruction operators.
However, there is another point of view which is more related to
our present understanding of quantum gravity.
It is based on the semi-classical behaviour of some dynamical variable
(say $a$). In physical circumstances in which $a$ behaves ``almost''
classically, two different schemes can be used to describe quantum processes
occuring to the other variables. First, one can work in the
background field approximation for $a$ (in which case, one has quantum fields
in a given gravitational background). Secondly, we can work in an extended
quantum framework and use WKB waves for describing $a$. Then, in
order for the two versions of transition amplitudes to coincide, 
the Wronskians of the WKB waves are univoquely fixed.
Therefore, in the next sections, the gravitational
waves shall have unit Wronskians, since we must recover the Schr\" odinger
equation in the limit in which $a$ can be described at 
the background field approximation.

\section{The adiabatic treatement in quantum cosmology}\label{wdw}

\subsection{When matter energy is a conserved quantity}\label{cons}

As a warming up, we first
analyse the WDW equation in the simple case 
when the matter energy is a constant of motion
(which is a stronger condition than being adiabatically conserved).
This implies that states with different matter energy are 
completely decorrelated.
In the next sections we shall drop this assumption and see
how the coupling to gravity introduces the notion of evolution
through interferences among neighbouring matter states 
characterized by adiabatically conserved energy.

We work in a closed Robertson-Walker universe
whose metric is given in eq. (\ref{1}).
Then, only one constraint survives corresponding to the
reparametrisation invariance $t=t(t')$. Classically, this constraint is
\be
{\cal H}= {-G^2 \pi_a^2 - a^2 + \Lambda a^4 \over 2 G a} + 
H_M(\phi,\Pi,a) =0
\label{HWDW}\ee
where $a$ is the scale factor now promoted to a dynamical variable,
 $\pi_a$ its conjugate momentum,
$H_M$ is the matter hamiltonian, with $\phi$ and $\Pi$ canonically
conjugate matter degrees of freedom. 
$\Lambda$ is the cosmological
constant. It shall play no other role than to participate to the
determination of $\pi_a$ in terms of $a$ and $H_M$.

For the matter energy to be a
conserved quantity $H_M$ must have the form 
\be
H_M(
\phi,\Pi ,a) = a^p h_M(\phi,\Pi)
\label{simple}
\ee
(or a sum of such terms). 
The simple  example we consider  
a very 
massive field at rest (i.e. $k=0$).
Then
$H_M= h_M = (\Pi^2 + M^2  \phi^2)/2$ 
where $M$ is the mass of the particles, see \cite{wdwgf}. 
Here $\Pi$ and
$\phi$ are rescalled by a factor $a^{3/2}$ with respect to the fields
defined in section 2.3. This rescalling replaces the adiabatically invariant
states characterized by instantaneous occupation numbers
given in eq. (\ref{decomposeas}), by truly
invariant states characterized by fixed occupation number.

In quantum cosmology, the physical states $\Xi(a, \phi)$ are 
solutions of the WDW constraint 
\be
{\cal H}\ \Xi (a, \phi)= \left[ G^2 \partial_a^2 - a^2 + \Lambda a^4 + 2Ga 
H_M( \phi, i\partial_\phi, a)
\right]\Xi (a, \phi) = 0
\label{WDW}\ee
(We neglect any ambiguity due to ordering problems. These would add
terms proportional to $G^2/a^2$ which could easily be taken into account.)
When, $H_M$ is of the form eq. (\ref{simple}),
 $h_M$ is independant of $a$ and commutes with 
 ${\cal H}$.
We can thus work in a
basis of eigenstates of $h_M$
\ba
h_M \ket{\psi_n} &=& E_n \ket{\psi_n}\nonumber\\
\scal{\psi_m}{\psi_n}&=& \delta_{m,n}\ea
Notice that these eigenstates coincide with
the stationary eigenstates of the Schrodinger equation
$i\partial_t \ket{\psi (t)} = h_M \ket{\psi (t)}$, see section 2.1.
Using these states, we can decompose $\Xi$ as
\be
\Xi (a, \phi) =\sum_n {\cal{C}}_n \Psi_n(a) \bk{\phi \vert \psi_n} \label{xi1}
\ee
Thus, since $\partial_a \ket{\psi_n} =0$, the different terms in eq. (\ref{xi1}) are completely 
uncorrelated.
The $n$-th wave function $\Psi_n(a)$, entangled to the
matter state $\ket{\psi_n}$, obeys 
\be
\left( -G^2 \partial_a^2 - a^2 + \Lambda a^4 + 
2G a E_n \right) \Psi_n(a) =0
\label{eqCn}
\ee
It corresponds to the equation for 
an oscillator whose time ($=a$) 
dependent frequency is given by 
\be
\omega^2(a)= \left( - a^2 + \Lambda a^4 + 
2G a E_n \right)/G\label{momentum}
\ee 
The solutions of this equation have been 
discussed in section 2.4. 

To make contact with former works\cite{BV}\cite{vil}\cite{tunnel}, 
we recall how a time parameter 
can be extracted from wave packets made out WKB solutions of 
eq. (\ref{eqCn}).
Following\cite{wdwpt,wdwin}, we shall then recall 
why interactions leading to transitions
must be added in order to give
physical substance to the expressions.

The WKB solution of eq. (\ref{eqCn}) with positive unit wronskian is 
\be
\chi_n(a) = { e^{-i \int^a\! da\  p _n(a)} \over \sqrt{2 p _n(a)}}
\label{WKBcos}\ee
where the momentum $p_n(a)$ is the solution of eq. (\ref{momentum})
with $H_M = E_n$, hence given by
\be
p(a, E_n) = p_n(a) = {1 \over G} \sqrt{-a^2 + \Lambda a^4 + 2 G a E_n}
\label{WKBcos2}\ee 
These WKB solutions correspond to expanding 
universes. Their complex conjugate describe contracting universes.

Consider now a superposition, eq. (\ref{xi1}), made out of WKB solutions with 
positive Wronskian only.
Lets further assume that the coefficients ${\cal{C}}_n$
form a well peaked enveloppe centered around
the mean value $\bar n$ such that the phase shifts
appearing in the sum can be evaluated to first order in $n- \bar n $
only and that the dependance in $n- \bar n$ of the prefactors
$p_n(a)^{-1/2}$ be neglected, see subsection 3.3
for more details. Then eq. (\ref{xi1}) can be factorized as
\ba
\Xi (a, \phi) =
{e^{-i \int^a \! da p_{\bar n}(a)} \over \sqrt{2 p_{\bar n}(a)}} \
\left[
 \sum_n  \ {\cal{C}}_n \  e^{-i(E_n - E_{\bar n})
\; \partial_{E}\! \int^a\! da\;  p(a,E) } \; \bk{\phi \vert \psi_n} \right]
\ea
The factor of $E_n - E_{\bar n}$ 
\be
 \partial_{E} \int^a\! da \; p(a , E)\vert_{E = E_{\bar n}} = t_{\bar n}(a) 
\label{time}
\ee
is the lapse of proper time evaluated in the
classical geometry driven by the mean matter energy
$E_{\bar n}$\cite{ontime},
in virtue of Hamilton-Jacobi equations.
%

Thus, well peaked wave packets in quantum cosmology factorize
into a gravitational piece characterized by the mean matter energy and 
and a linear superposition of matter states multiplied
by the conventional phase factors
parametrised by the mean time $t_{\bar n}(a)$.
However, in this model
characterized by constants of motion, this quantum superposition
is physically empty 
since there is nothing to probe
the evolution, i.e. to give physical meaning to the 
phase factors which arise precisely from the comparison of neighbouring 
gravitational waves entangled to the corresponding matter states. 
Therefore, neither a physical interpretation can be given to the 
coefficients ${\cal{C}}_n$ nor can a principle be invoqued 
to normalise them. 

On the contrary when
the matter hamiltonian no longer commutes with the WDW
constraint (i.e. which now depends on $a$) the interactions among matter
states gives a precise physical meaning to
both the phase $\Delta E \int \! da \partial_E p(a,E) = \Delta E t $ and 
to the coefficients
${\cal{C}}_n$ whose normalisation will be univoquely fixed.
 Indeed, $\Delta E t $ is the phase which 
accumulates between successive transitions
and therefore which governs their rate.
And the coefficients
${\cal{C}}_n$ obey a Schr\" odinger equation only
if the Wronskians of the corresponding
gravitational waves $\chi_n(a)$ are all the same, 
see the end of section 2.4.

\subsection{Non adiabatic transitions in quantum cosmology}\label{nonadQC}

In this subsection, the matter hamiltonian depends explicitly on
$a$. Examples of such Hamiltonians where analysed in section 2.2 and
2.3. They will be further discussed in section 4. 
The coefficients
${\cal{C}}_n$ then depend on $a$.
To obtain the equation which govern their evolution,
we must extend the usual adiabatic procedure given in section 2.1.
Indeed, because of the second order character of the WDW equation,
a {\it double} adiabatic development is now required
to obtain a linear equation in $\partial_a$
for the ${\cal{C}}_n$, 
First, as in section 2.1, we diagonalise instantaneously
the matter hamiltonian
and secondly, we apply an adiabatic decomposition
of the type given in eq. (\ref{wkbexp})
to the gravitational wave functions entangled to the 
instantaneous matter eigenstates.
Moreover, 
in order to determine univoquely the coefficients
${\cal{C}}_n$, an additional mathematical condition must still
be supplied. We proceed formaly by imposing that $\Xi(a, \phi)$
satisfies an equation that generalyses eq. (\ref{OTHER}). 
The resulting equation
for the ${\cal{C}}_n$ 
describes
both non-adiabatic transitions between matter eigenstates and the
backscattering from expanding universes into
contracting ones.
The physical meaning of this equation
will be analyzed in the next sections.

Thus, first, we carry out an instantaneous 
diagonalisation, i.e. at fixed $a$, of $H_M$:
\ba 
H_M(a) \ket{\psi_n(a)}&=& E_n(a) \ket{\psi_n(a)}
\nonumber\\
\scal{\psi_m(a)}{\psi_n(a)}&=& \delta(m,n)
\ea
We emphasize once more that this diagonalisation does not
require the ``existence'' of a Schr\" odinger equation, although
the eigenstates $\ket{\psi_n(a)}$ and the eigenvalues $E_n(a)$ 
coincide with ones that one would have obtained if one 
had started, as in section 2.1, from the Schr\" odinger equation whose r.h.s. 
contains the hamiltonian given by $H_M(a(t))$.

Using these instantaneous eigenstates,
$\Xi(a, \phi)$, solution of the Wheeler-DeWitt equation
(\ref{WDW})
can be decomposed, as in eq. (\ref{xi1}) but without any lost
of generality, as
\be
\bk{\phi \vert \Xi(a)} = \Xi(a, \phi) = \sum_n \varphi_n(a) \bk{\phi \vert \psi_n(a)}.
\label{decomp1}
\ee
The difference with eq. (\ref{xi1}) is that the waves $\varphi_n(a)$ 
are no longer decorrelated since the instantaneous eigenstates
$\ket{\psi_n(a)}$ do now depend on $a$.
In section 3.4, the correlations among the $\varphi_n(a)$ will be related
to the non-diagonal matrix elements of $\partial_a \pi_a^2$, the 
kinetical term of gravity viewed as an operator acting in
the matter Fock states.

Up to now, no difference exists between the present development
and the adiabatic treatement of section 2.1. 
The difference arise from the fact that the WDW constraint is
second order in $\partial_a$. Therefore, to obtain a first order equation
for the coefficients ${\cal{C}}_n$, we must proceed 
to a second adiabatic development.
To this end, we express each $\varphi_n(a)$ in terms of the 
WKB waves $\chi_n(a)$, eqs. (\ref{WKBcos}, \ref{WKBcos2}) with
$E_n$ replaced by $E_n(a)$:
\be
\varphi_n(a) = {\cal{C}}_{n}(a)\ \chi_{n}(a) 
+  {\cal{D}}_{n}(a) \ \chi_{n}^*(a)
\label{decomp22}
\ee
In the limit where both the adiabatic approximation for the matter states
and the WKB approximation for gravity are valid, 
the coefficients ${\cal{C}}_{n}$ and ${\cal{D}}_{n}$ are 
constants. In that case, we recover the situation presented in section 3.1
wherein there is no correlations among the coefficients 
 ${\cal{C}}_{n}$ and ${\cal{D}}_{n}$.

To describe these correlations in the general case, we first need to
complete the definition of ${\cal{C}}_{n}$ and ${\cal{D}}_{n}$.
We proceed mathematically. Notice however that this
mathematical choice has deep physical consequences
since we shall attribute a physical 
interpretation to the ${\cal{C}}_{n}$.
We generalyse eq. (\ref{OTHER}) and require 
\be
\bra{\psi_n(a)} i\!\! \stackrel{\rightarrow} \partial_a \!\ket{ \Xi(a) }= p_n(a)
\left[ {\cal{C}}_{n}(a)\ \chi_{n}(a) -  {\cal{D}}_{n}(a)\ \chi^*_{n}(a)
\right]
\label{AUXWKB}\ee  
The reason for this choice is that it 
ensures that the total current carried by  $\Xi(a, \phi)$
contains no terms proportional to $\partial_a {\cal{C}}_{n}(a)$ or $
\partial_a {\cal{D}}_{n}(a)$. Indeed, one has
\be
\bk {\Xi(a) \vert i 
\lrpartial{a} \vert \Xi (a)} = \int d\phi \ \Xi^*(a, \phi) i 
\lrpartial{a} \Xi (a, \phi) = \sum_n \vert {\cal{C}}_{n}(a) \vert^2 - 
\vert {\cal{D}}_{n}(a) \vert^2 = C
\label{CONSCURR}\ee
Notice that the absence of relative factors in the above sum
directly follows from our choice of identical (unit) Wronskians.
The same remark also applies to the next two equations.
We are now in position to determine the $a$-dependance of 
${\cal{C}}_{n}$ and ${\cal{D}}_{n}$.
Taking the derivative of eq. (\ref{decomp1}) 
and using eq. (\ref{AUXWKB}) yields 
\be
\partial_a {\cal{C}}_{n} \chi_{n} + \partial_a {\cal{D}}_{n} \chi_{n}^*
-{\partial_a p_n \over 2 p_n} ({\cal{C}}_{n} \chi_{n} +  {\cal{D}}_{n}\chi_{n}^*)
+\sum_m\scal{\psi_n}{\partial_a \psi_m}
 ({\cal{C}}_{m} \chi_{m} + {\cal{D}} _{m}\chi_{m}^*)
=0\label{AUX2}\ee
Then taking the derivative of eq. (\ref{AUXWKB}), 
inserting it in the WDW equation, 
and using eq. (\ref{AUX2}) to eliminate either $\partial_a {\cal{C}}_{n}$ 
or $\partial_a {\cal{D}}_{n}$ yields
\ba
\partial_a {\cal{C}}_{n} &=& \sum_{m \neq n}
\scal{\partial_a \psi_m}{\psi_n}
{p_n + p_m\over 2\sqrt{p_n p_m}}
e^{-i \int ^a (p_n - p_m) da }
 {\cal{C}}_{m}\nonumber\\
& & + \sum_m \scal{\partial_a \psi_m}{\psi_n}
{p_n - p_m\over 2\sqrt{p_n p_m}}
e^{-i \int ^a (p_n + p_m) da }
{\cal{D}}_{m}\nonumber\\
& &+{\partial_a p_n \over 2 p_n} 
e^{-2i \int ^a p_n da } {\cal{D}}_{n} 
\label{central}\ea
and the same equation with ${\cal{C}}_{n} \leftrightarrow {\cal{D}}_{n}$,
$i \leftrightarrow -i$.

This is our central equation. We emphasize that
it is equivalent to the original
WDW equation (\ref{HWDW}).
Its physical meaning is
investigated in the next sections.

\subsection{Time evolution and unitary evolution}

To begin the analysis 
of eq. (\ref{central}),
we assume in this section that the coupling between expanding
and contracting universes may be neglected.
In the next section, we shall take this unusual quantum
coupling into account.

Thus, we drop the second and third term 
on the r.h.s. of
eq. (\ref{central}). This yields
\be
\partial_a {\cal{C}}_{n} = \sum_{m \neq n} \scal{\partial_a \psi_m}{\psi_n}
{p_n + p_m\over 2\sqrt{p_n p_m}}
\; e^{-i \int ^a (p_n - p_m) da } \; {\cal{C}}_{m}
\label{centralP}
\ee
In what follows, we shall proceed to the interpretation of this
equation in three steps by
progessively relaxing  restrictions imposed on the coefficients
${\cal{C}}_{n}$.

When comparing eq. (\ref{centralP}) to eq. (\ref{eqtcn2}), we see that the main
difference is that, in  eq. (\ref{centralP}), there are additional
dependances in the matter states through the momentum
of gravity $p_n(a)$.
 Therefore, in a first step, 
we assume that the ${\cal{C}}_{n}$ form a well peaked wave packet centered 
around ${\cal{C}}_{\bar n}$ such
that 
a first order expansion in $n -\bar n$
is a valid approximation.
Then, using eq. (\ref{time}), we obtain
\be
\partial_{a} {\cal{C}}_{n} = \sum_{m \neq n} \scal{\partial_{a} \psi_m}{\psi_n}
\; e^{-i \int ^{t_{\bar n}(a)} (E_n - E_m) dt } \;
{\cal{C}}_{m}
\label{centralS}
\ee
This equation is identical to eq. (\ref{eqtcn2}). 
More precisely, through $t_{\bar n}(a)$, the proper
time evaluated in the mean universe,
we recover the Schrodinger equation governing
non adiabatic transitions among instantaneous matter states
{\underline {if}} one identifies the coefficients ${\cal{C}}_{n}$
with the
probability amplitudes $c_n$ to find the matter states at $t_{\bar n}(a)$.
We emphasize the a posteriori character of this identification.
Indeed, it is based on the comparison of eq. (\ref{eqtcn2})
with the resulting simplified equation governing the ${\cal{C}}_{n}$.
We recall that the definition of the coefficients ${\cal{C}}_{n}$ 
required three equations, namely eqs. (\ref{decomp1}, 
\ref{decomp22}, \ref{AUXWKB}).
Had we made other choices, we would not have obtained this
straitforward identification.
Moreover, two approximations were also necessary:
first that the coupling between expanding
and contracting universes be neglected and secondly that
the ${\cal{C}}_{n}$ form a well peaked distribution.
For this later condition to be met, we recall that 
the universe be macroscopic, see \cite{BV,vil,wdwpt}.
The generic character of the recovery 
of eq. (\ref{eqtcn2}) from eq. (\ref{central}) through these
approximations should be re-emphasized. This whole 
scenario arises whenever one considers how to 
reach the background field approximation from a  
more quantized framework, see the end of section 2.4 
and \cite{rec,su}.

As in the cases treated in these references, 
eq. (\ref{centralP}) contains higher order terms in $n -\bar n$
which define the deviations from the Schr\" odinger equation
when the ${\cal{C}}_n$ form a well defined wave
packet. This condition must be met in order to have
 physically meaningfull expansions around mean values. 
Eq. (\ref{centralP}) 
differs from eq. (\ref{centralS}) 
through both the complex phase 
and the prefactors.
Both give rise to gravitational corrections 
since the differences arise
from the quantum backreaction expressed here by the
dependance in $n$ of $p_n(a)$. 
To analyse these corrections we define new coefficients 
$\tilde  {\cal{C}}_{n}=  {\cal{C}}_{n} e^{-i \int ^a p_n da }$
which allow to rewrite eq. (\ref{centralP}) as
\ba
i \partial_a \tilde {\cal{C}}_{n} = 
\sum_m \left( \delta_{mn} p_n + i \scal{\partial_a \psi_m}{\psi_n}
{p_n + p_m\over 2\sqrt{p_n p_m}}
\right)  \tilde {\cal{C}}_{m} = \sum_m H^{eff}_{nm}(a) \tilde {\cal{C}}_m
\label{heff}\ea
where we have defined, as in eq. (\ref{eqcn}), 
an effective hamiltonian giving the
evolution of $\tilde {\cal{C}}_{n}$.
We now expand the $p_n(a)$ 
around the mean energy $\bar E(a)$.
 to obtain
\ba
H^{eff}_{nm}(a) &=&
\delta_{mn} p_{\bar n} (a) \nonumber\\
& & + \delta_{mn} (E_n - \bar E) \partial_{E} p_{\bar n} + 
i\scal{\partial_a \psi_m}{\psi_n} \nonumber\\
& & + \delta_{mn} { 1\over 2}  (E_n - \bar E)^2   \partial^2_{\bar E} 
p_{\bar n} 
+ i \scal{\partial_a \psi_m}{\psi_n} 
{(\partial_ {\bar E} p_{\bar n})^ 2 \over 8 p_{\bar n}^2 } ( E_n - E_m
)^2
\nonumber\\
&& + O(\Delta E^3)
\label{heffcorr}\ea
The first line 
determines the overall phase of the $\tilde {\cal{C}}_{n}$
and encodes no physics.
The second line contains the linear deviation in energy
and gives the Schrodinger 
equation in the mean geometry, eq. (\ref{centralS}). 
The third line 
contains quadratic terms which furnish the gravitational
corrections to the energy levels and the instantaneous eigenstates.
It consists of two terms. 
The first term arises because the energy of matter modifies the
propagation of gravity, and hence the phase accumulated by the wavefunction.
It can be expressed as a shift of the energy levels
\be
\Delta^{Grav} (E_n) = { 1\over 2} {\partial^2_{\bar E} p 
\over \partial_{\bar E} p} (E_n - \bar E)^2
\simeq -{\Delta E^2 \over 4 \bar E}
\label{gravcorr}
\ee
where in the second equality we have evaluated the correction
for a matter dominated universe in which $p_{\bar E} \simeq \sqrt{ a
\bar E/G}$. (Notice that $\partial^2_{\bar E} p $ plays a role
similar to the specific heat of a large system, see App. B of
\cite{wdwgf}.) The second term of the third line in eq. (\ref{heffcorr}), 
is the first gravitational correction to the 
non adiabatic coupling. 
By diagonalising $H^{eff}$ one can calculate how it 
modifies the energy levels.
This shift will be much smaller than $\Delta^{Grav} (E_n)$ of eq.
(\ref{gravcorr}) because it
carries an additional factor $1/p$. 

Because of these corrections, there will be a maximum time 
$t_{max}= O( \bar E^2/(E_n - \bar E)^2) $ after which errors
in the phases of the ${\cal{C}}_n(a)$ 
are of order 1.  Then the evolution of the
${\cal{C}}_n(a)$ determined by the WDW
equation no longer coincides with the one of the $c_n(t)$, solutions of the
Schr\" odinger equation.
The shift in the energy levels will also modify the non
adiabatic transition amplitudes through the modification of the
saddle point amplitude. 
At this point, it is to be emphasized that 
the calculation of these effects in specific cases is best
carried out directly through the effective Hamiltonian eq. (\ref{heff}),
or with eq. (\ref{centralP}).
Indeed the expansion eq. (\ref{heffcorr}) depends explicitly on the mean energy
$\bar E$ determining the background, whereas the transition amplitudes are
independent of $\bar E$. Therefore quantum cosmology tells us that the
(traditional) understanding of physics as describing matter processes
in a given gravitational  background should be considered as a
convenient 
approximation
with no dynamical justification. This point will be illustrated
in section 4 by some examples and further clarified in what follows.

We note indeed that
the validity of
eq. (\ref{centralP}) puts no restriction on the ${\cal{C}}_n$ since 
it requires only the decoupling between expanding and contracting universes.
This is sufficient to guarantee that 
$\sum_n \vert {\cal{C}}_{n} (a) \vert^2 = constant$.
Therefore one must still interpretate ${\cal{C}}_{n}(a)$ as the 
amplitude of probability to find matter in the n-th state at $a$, even
if they do not form a well defined wave packet\cite{wdwin}. 
In this case of arbitrary distribution of ${\cal{C}}_n$, it is completely
 meaningless to insist in using a mean background 
as a reference to parametrize
transitions. Thus this raises the interesting question
of whether one should require at all that the ${\cal{C}}_{n}$
form a well peaked wave packet in order to recover
the notion of a temporal evolution.
To this end 
we note that if one assumes that the matter
hamiltonian is described by a field theory with a cutoff 
(at the Planck
energy), non adiabatic transitions never occurs between states whose
energy differs by more than one Planck mass. Thus components of $\Xi$ with
very different matter energies will essentially never interact.
This offers the possibility that various components of the 
total waves decohere and evolve with their own time.
We will not attempt further discussion of this issue here.
Note however that this problem possesses an analogue in the emergence of a
temporal description of electronic transitions in atomic collisions in
which the atoms are described by broad wave packets, see the introduction.

\subsection{Backscattering effects}\label{universe}

We now turn to the coupling between expanding and contracting
universes, that is the coupling  
between
the coefficients ${\cal{C}}_{n}$ and ${\cal{D}}_{n}$ in eq. (\ref{central}). 
This coupling is inevitable because the WDW equation is second order
in $\partial_a$. It also occurs in the simple harmonic oscillator
problem discussed in section 2.4. However in the present case the
situation is more complicated because there are backscatterings which
keep the matter state unchanged, i.e. the third term in eq. (\ref{central}),
and backscatterings which change the matter 
state, i.e. the second term in eq. (\ref{central}). 
All of these couplings however have the same origin: 
they can all be expressed in terms of matrix elements of 
\be
\pi_a^2(a;\phi,\Pi) = (\Lambda a^4 - a^2)I + 2 a H_M(a;\phi,\Pi)
\ee
viewed as an operator parametrised by $a$ and acting on matter states
 (here $I$ is the identity operator on these states).
Indeed, using eq. (\ref{dHdt}), eq. (\ref{central}) can be written as
\ba
i \partial_a {\cal{C}}_{n} &=&
\sum_{m\neq n}
i\bra{\psi_m} \partial_a \pi_a^2\ket{\psi_n} 
{ 1 \over 2 (p_n - p_m) \sqrt{ p_n p_m}} {\cal{C}}_m\nonumber\\
&&+
\sum_m i \bra{\psi_m} \partial_a \pi_a^2\ket{\psi_n} 
{ 1 \over 2 (p_n + p_m) \sqrt{ p_n p_m}} {\cal{D}}_m
\label{newrewr}
\ea
The first term in this equation corresponds to
 the first term in eq. (\ref{central}), and the second term regroups the
two last terms of eq. (\ref{central}). 
Therefore, the operator $\partial_a \pi_a^2$
generalyses the operator $\partial_t H(t)$ in the Schr\" odinger case, see
eq. (\ref{dHdt}), in that it allows this very simple rewriting of the
WDW equation.

The coupling between forward and backward propagating universes 
described by eq. (\ref{newrewr}) gives
rise to new effects. First it predicts the generation of 
backward propagating waves, i.e. ${\cal{D}}_n \neq 0$, 
 from initialy purely forward waves and the simultaneous
increase the norm of the ``forward'' coefficients ${\cal{C}}_n$ so as to ensure
conservation of the Wronskian eq. (\ref{CONSCURR}). This is strictly the
analogue of the Klein paradox in relativistic quantum field theory.
Indeed, no positive local\footnote{
See \cite{embacher} for recent attemps to build non-local
positive definite conserved quantity.} conserved quantity 
can be either constructed from the solutions of the Wheeler-DeWitt,
contrary to the situation in quantum mechanics. It is possible that the
resolution of this problem requires that one proceeds to third
quantization wherein $\Xi(a)$ is interpreted as an Heisenberg operator
acting on universe Fock states.

Secondly, eq. (\ref{newrewr}) stipulates that there are quantum 
interferences among matter states associated with the expanding
and contracting solutions. This means that the knowledge of the coefficients
${\cal{D}}_n(a)$ is required to determine the evolution of the ${\cal{C}}_n(a)$
at other values of $a$. Thus the set composed of the ${\cal{C}}_n$ only form
an open system.

In a future publication we hope to return to these aspects of quantum
cosmology.

\section{Applications of the formalism}\label{Appl}

In this section we return to the examples of non adiabatic transitions
of section 2. We shall now treat them in the context of quantum
cosmology when gravity is described in the adiabatic approximation
hence with the amplitudes of matter transitions 
described by eq. (\ref{centralP}). 
The essential difference is that whereas in section 2 it was merely
convenient to parametrise the evolution by $a$, see eqs. (\ref{eqtcn}, \ref{eqtcn2}),
 here it is mandatory
to do so. This is particularly clear when calculating gravitational
corrections to the transition amplitudes, since these corrections
depend only on the structure of the energy levels $E_n(a)$, and are
independent of the mean energy $\bar E$ and the mean time $\bar t$ 
defining the reference background. 

However before starting we must address the following problem.
In the examples treated in section 2,
 the particles have non vanishing momentum
which will backreact
onto gravity and breack the symmetry of minisuperspace. 
The way around this difficulty is to note\cite{HalH,wdwpt}
that if the momentum of the particles is
small, then one can treat the corresponding metric fluctuations
perturbatively.
Thus one expands both of the spatial metric $h_{ij}$ and the lapse and
shift in a
fourier series, and keeps in the
Einstein-Hilbert-matter action only quadratic terms 
with $k\neq 0$. The Hamiltonian then becomes 
\be
\int d^3 x N^\mu H_\mu
= N^0(k=0) H_0(k=0) + \sum_{k\neq 0} N^\mu(k)H_\mu(-k)
\ee
 All the new
constraints $H_\mu (k)$  which are multiplied by $N^\mu(k\neq 0)$
are linear in the gravitational 
perturbation $h_{ij}(k)$ and simply serve to eliminate the non
propagating parts of $h_{ij}$. The only non trivial constraint is
$H_0(k=0)$. Thus to this order, one has recovered a minisuperspace
model, but which contains matter and linearised gravitons with non
vanishing momenta.
Since they are not coupled, one can treat
them seperatly. We shall for simplicity only consider the
matter part of the Hamiltonian. 

We first consider the recovery of the Golden Rule in quantum cosmology.
The instantaneous eigenstates $\ket{\psi_\pm}$ and eigenenergies $E_\pm$
where already obtained in section 2.2. Therefore the
effective hamiltonian eq. (\ref{heff}) 
for the coefficients $\tilde {\cal{ C}}_\pm$ multiplying 
$\ket{\psi_\pm}$ is 
\be
H^{eff}(a) =
\left( 
\begin{array}{cc}
p_+(a) & i \scal{\partial_a \psi_-}{\psi_+} {p_+ + p_- \over 2
  \sqrt{p_+ p_-}}\\
-i \scal{\partial_a \psi_-}{\psi_+}{p_+ + p_- \over 2
  \sqrt{p_+ p_-}} &p_-(a)
\end{array}
\right)
\simeq \left( 
\begin{array}{cc}
p_+(a) & i \scal{\partial_a \psi_-}{\psi_+}\\
-i \scal{\partial_a \psi_-}{\psi_+}&p_-(a)
\end{array}
\right)\label{Heffatom}
\ee
where in the second equality we have neglected the gravitational
corrections to the non adiabatic couplings.
Here the momenta of gravity $p_{\pm}$ characterizing the gravitaional 
waves entangled to the matter states $\ket{\psi_\pm}$ are
\ba
p_\pm &=& \sqrt{2 G a ( E_T + E_\pm(a))}\nonumber\\
&=& 
\sqrt{2 G a ( E_T + {M + m +k/a  \over 2}
\pm \sqrt{
(M - m -k/a)^2/4 + g^2 / a^2 k}
)}
\label{newps}
\ea
We have chosen to work in a flat 
matter dominated universe with conserved matter
energy $E_T$.
Since  eq. (\ref{Heffatom}) has exaclty 
the same structure as eq. (\ref{eqcn}),
we can calculate the non adiabatic transition amplitude as in section 2. 
We recall that the probability to make a non
adiabatic transition 
is the probability for the excited atom not to decay.
In the saddle point method
it is given by
\ba
P_{no\ transition}(k) = e^{- 2 Im \int^{a^*}\!da \; \left[p_+(a) - p_- (a)\right] }
\ea
where $a^*$ is the complex value of $a$ where the momenta of the
universe are equal 
\be
p_+(a^*) - p_-(a^*)= 0
\ee
This gravitational resonnance condition leads back to 
the background field resonnance condition $E_+(a^*) - E_-(a^*)= 0$, 
since $p_{\pm}(a) = p( a, E_{\pm})$, see eq. (\ref{newps}).
Thus $a^*$ is unafected by the dynamical character of gravitys 
and still given by eq. (\ref{compa}).

The integral which yields the transition
probability is simply
\ba
I_k &=& Im \int_{a_0}^{a_+^*}\!da \; \left[p_+(a) - p_-(a)\right]
\ea
In order to estimate this integral, we first expand the
integrant around the average total energy
$\bar E = E_T + (E_+ + E_-)/2$.
The important word in the previous sentence is ``total''.
This is why this expansion is physically meaningfull:
the mass-energy of the matter content of the universe
which is not involved in the process determines the ``inertia'' 
of gravity. Indeed, we obtain
\ba I_k
&=&Im \int_{a_0}^{a^*}\!da \; \partial_E p(a,E)\vert_{E = \bar E} \;
\left[ E_+(a) - E_-(a) \right] + O\left({(E_+ - E_-)^3 \over E_T^3 }\right)
\nonumber
\\
&=&  Im \int_{a_0}^{a^*}\!da \; {d\bar t \over da} (a) \;
\left[ E_+(a) - E_-(a) \right]
 + O\left({(E_+ - E_-)^3 \over E_T^3 }\right)\label{Ik1}
\ea
Note that this is still not strictly equivalent to the 
background field approximation since the ``time'' parameter $\bar t$ 
determined by 
\be
\partial_E p(a,E)\vert_{E = \bar E} = { d \bar t \over da}
\ee
depends {\it both} on the
initial state of energy $E_T +M$ and on the final state of energy $E_T
+ m + k/a$. Thus successive transitions, corresponding to successive
$k$, are described by (slightly) different time parameters.
A single classical time parameter could be chosen to consist of the atom in
its excited state of energy $M$, call it $\tilde t(a)$. This is particularly appropriate in
the present case because we are considering the probability for the
atom to stay in its excited state.
Then eq. (\ref{Ik1}) becomes
\ba
I_k 
&=&Im \int_{a_0}^{a^*}da 
{d\tilde t \over da} (a)\; \left[ E_+(a) - E_-(a) \right]
(1 - { \Delta M \over 2 (E_T + M)} {(a-a_0)\over a_0} + O((a-a_0)^2)) \nonumber\\
& &\quad
+ O\left({(E_+
- E_-)^3 \over E_T^3 }\right)
\ea
The first term corresponds to eq. (\ref{integralpos}) 
evaluated in the background
driven by $E_T + M$. The second term 
is the first gravitational correction
to the decay probability. It leads to a modification of the decay rate
of the order of $\Delta M/ E_T$. 
Note that the correction decreases with increasing
$a_0=k/\Delta M$. Indeed as the universe expands it is larger and larger and
the background approximation becomes better and better.

We now apply the same techniques to 
the calculation of 
pair creation amplitudes in quantum cosmology.
As in section 2.3 we consider only the
transition from the vacuum state to the state containing one pair of
momentum $k,-k$.
The only relevant quantities are the momenta of gravity
associated with these two states. They are given by 
\ba
p_0(a) &=& \sqrt{\La a^4 - a^2 }\nonumber\\
p_{k,-k}(a) &=& \sqrt{\La a^4 - a^2  + 4 G a \sqrt{m^2 + k^2/a^2}}
\ea
Then, in  the saddle point
calculation, the amplitude to produce a pair is simply given by 
\be
\vert {\cal A}_{pair}\vert^2 
= e^{-2 Im \int^{a^*}\!da \;
 \left[p_0(a) - p_{k,-k}(a)\right]}
\label{pairW}
\ee
The saddle point $a^*$ of eq. (\ref{pairW}) is again determined by the
equality of the two mgravitaional momenta: $p_0(a^*) - p_{k,-k}(a^*)=0$. 
Thus  $a^*$ is still given by $a^*= i k/m$, which is
the result obtained in the background field approximation
eq. (\ref{tstar}).

To estimate the integral of eq. (\ref{pairW}) one can expand the
integrant around empty de Sitter space, by exploiting the inertial
delivered by the cosmological constant. However, it is 
more interesting to note that this expansion around a background
is neither physically intrinsic nor mathematically necessary.
Indeed,  the integral of eq. (\ref{pairW}) is easily evaluated by
deforming the integration over $a$: 
by first integrating along the real axis to $a = +\infty$,
then along the circle at infinity from 
$a= +\infty$ to $a= +\infty e^{i \pi/2}$
and fianlly along the imaginary axis from $a = i \infty$ to $a = a^*= ik/m$.
The first piece of the integral, being real, does not contribute to the 
pair production probability, the second piece yields the background field
result, eq. (\ref{paircreat2}),
and the third piece yields the
gravitational corrections to the pair production probability. One
finds that for large $k$, corresponding to productions occuring far
from the classical turning point $a= 1/\sqrt{\La}$, the probability is
\be
\vert {\cal A}_{pair}\vert^2 \simeq e^{{-2 \pi m \over \La} + O( G
  m^5/k^3 \La^{3/2})}
\ee
Thus in minisuperspace, the gravitational corrections to pair creation
decrease as the time (which is related to $k$ by the real part of
eq. (\ref{tstar})) of production increases. This is natural since at
late times the inertia delivered by the cosmological constant
is huge, and thus, accordingly, the background field approximation better and better. 
It should however be noted that there probably are {\it local} gravitational
corrections to pair creation 
which survive even at late times. In particular in de
Sitter space these should be related to changes of the area of the
horizon\cite{thermaldeSitter}. Similar corrections should also obtain
for the radiative transitions discussed in the previous paragraph.

\section {Conclusion}

In this article we have developped a formalism for analysing the
Wheeler-DeWitt
equation based on a double adiabatic developpement. The matter part of
the
wave function
was expanded in terms of the adiabatic matter states and the
gravitational part 
in terms of the adiabatic (WKB) waves.
The equation for the
coefficients of this develloppement is the main result of this paper.
Because of its generality this formalism is usefull to
study conceptual problems such as the recovery of a temporal evolution
for matter, and the identification of the probability amplitude to be
in a given matter state. 
We have applied it to two examples: radiative transitions and
particle creation, and we have compared how these processes are
described in the background field approximation and in quantum cosmology.

\vskip 1. truecm

{\bf Acknowledgements}\\
S. M. would like to acknowledge partial support 
by  grant 614/95 of the
Israel Science Foundation.
\\

\end{document}